# PROPERTIES OF ISOSCALAR GIANT MULTIPOLE RESONANCES IN MEDIUM-HEAVY CLOSED-SHELL NUCLEI: A SEMIMICROSCOPIC DESCRIPTION

M.L. Gorelik [1], S. Shlomo [2,3], B.A. Tulupov [4], M.H. Urin [5]

1) Moscow Economic School, Moscow 123022, Russia

2) Cyclotron Institute, Texas A&M University, College Station, TX 77843, USA

3) Department of Elementary Particles and Astrophysics, the Weizmann Institute of Science, Rehovot 76100, Israel

4) Institute for Nuclear Research, RAS, Moscow 117312, Russia

5) National Research Nuclear University "MEPhI", Moscow 115409, Russia

The semimicroscopic particle-hole dispersive optical model (PHDOM) is implemented to describe main properties of Isoscalar Giant Multipole Resonances (up to $L = 3$) in medium-heavy closed-shell nuclei. The main properties are characterized by the strength distribution, transition density, partial and total probabilities of direct one-nucleon decay. Calculation results obtained for the $^{208}$Pb nucleus are compared with available experimental data.

## I. INTRODUCTION

A detailed description of any giant resonance (GR) [1, 2] includes the following characteristics: (i) strength distribution for a large excitation-energy interval; (ii) energy-dependent double transition density (which depends only on nuclear structure) and projected one-body transition density (associated with a given one-body probing operator), and; (iii) partial and total probabilities of direct one-nucleon decay. In order to get such a description, within theoretical approaches, the main relaxation modes of high-energy particle-hole-type ((p-h)-type) states associated with GRs should be together taken into account. These relaxation modes include Landau damping, coupling of the mentioned (p-h)-type states to the single-particle continuum, and the coupling to many-quasiparticle configurations (the spreading effect). The recently developed particle-hole dispersive optical model (PHDOM) [3] accounts for the above-described relaxation modes. PHDOM is a microscopically-based extension of the standard [4] and non-standard [5] versions of the continuum-random-phase approximation (cRPA) to taking the spreading effect into account. Within the PHDOM, which is a semimicroscopic model, Landau damping and coupling to the continuum are considered microscopically (in terms of a mean field and p-h interaction), while the spreading effect is treated phenomenologically (in terms of a properly parameterized energy-averaged p-h self-energy term). The imaginary part of this term determines the real part via a dispersive relationship. The current PHDOM version has been implemented to describe the simplest photo-nuclear reactions accompanied by excitations of the Isovector Giant Dipole Resonance [6], the main properties of the Isoscalar Giant Monopole Resonance in $^{208}$Pb [7, 8], properties of Isobaric Analog Resonance and its overtone in the mentioned parent nucleus [9]. Week violations of model unitarity (caused by the method of taking the spreading effect into account) and methods of unitarity restoration have been studied in Ref. [8].

In the present work, we implement the PHDOM current version to describe the main properties of the Isoscalar Giant Multipole Resonances (ISGMPRs) (up to $L = 3$) together with the overtones of Isoscalar Giant Monopole and Quadrupole Resonances. A cRPA-based description of isoscalar bound states, including the $1^-$ spurious state related to center-of-mass motion, is also taken into consideration. Calculation results obtained for the $^{208}$Pb nucleus are compared with available experimental data.

This work is motivated by the following possibilities: 1) to check abilities of the model in describing the strength distribution of isoscalar non-spin-flip GRs with taking into account the isobaric symmetry and translation invariance of the model Hamiltonian (integral characteristics of the distribution are compared with related results obtained within the microscopic RPA-based approach of self-consistent Hartree-Fock (HF) using Skyrme-type forces (SF) [2, 10]); 2) to get a microscopic input (the projected energy-dependent one-body transition density [7]) for an analysis (based on the Distorted Wave Born Approximation) of (α,α')-scattering cross sections of ISGMPRs excitation [11, 12] (as a rule, the quasi-classical collective-model (energy-independent) transition densities are used in such an analysis (see, e.g., Ref. [13, 14])), and; 3) to realize the unique ability of the model in describing the branching ratios for direct one-nucleon decay of ISGMPRs. The above-listed points allow us to point out that the use of PHDOM opens new possibilities in describing properties of non-spin-flip isoscalar GRs when compared with widely used (HF-based RPA +SF) approaches (see, e. g., the review paper of Ref. [15]). Some preliminary results of the present study are given in Ref. [16].

In Section II, we give the PHDOM basic equations and relations for describing ISGMPRs. Choice of model parameters, calculation results and a comparison with available experimental data are presented in Section III. Section IV contains a discussion of the results and conclusive remarks.

## II. BASIC EQUATIONS AND RELATIONS

Since the PHDOM is an extension of the cRPA versions to taking the spreading effect into account, most of the equations and relations of these approaches are similar. So, the basic equation for implementing the PHDOM is the Bethe-Goldstone-type equation for the energy-averaged (local) p-h Green function. In applying to the description of ISGMPRs in spherical nuclei, this Green function (or p-h propagator) can be expanded in spherical harmonics:

$$\tilde{A}(\vec{r},\vec{r}',\omega) = \sum_{LM}(rr')^{-2}\tilde{A}_L(r,r',\omega)\,Y_{LM}(\vec{n})Y^*_{LM}(\vec{n}')\,, \qquad (1)$$

where $\omega$ is the excitation energy. If the p-h interaction responsible for long-range correlations in the isoscalar and isovector non-spin-flip channels is taken in the form of Landau-Migdal forces

$$F(\vec{r}_1,\vec{r}_2) \to (F(r_1) + F'\vec{\tau}_1\vec{\tau}_2)\delta(\vec{r}_1 - \vec{r}_2) , \qquad (2)$$

one gets the equation for the p-h propagator radial components:

$$\tilde{A}_L(r,r',\omega) = A_L(r,r',\omega) + \int A_L(r,r_1,\omega)F(r_1)\tilde{A}_L(r_1,r',\omega)\,dr_1/r_1^2 . \qquad (3)$$

Here, the radial component $(rr')^{-2}A_L(r,r',\omega)$ of the "free" p-h propagator, which relates to the model of non-interacting and independently damping p-h excitations, is the key quantity in the PHDOM-based description of corresponding ISGMPR. The explicit expression for this quantity is discussed below. As a comment to Eq. (3), we note that weak mixing of $L \neq 0$ ISGMPRs with respective Isoscalar Giant Spin-Multipole Resonances due to the mean-field spin-orbit part is neglected.

The p-h propagator of Eq. (3) determines the corresponding component of the energy-averaged double transition density $\rho(\vec{r},\vec{r}',\omega) = \sum_{LM}(rr')^{-2}\,\rho_L(r,r',\omega)Y_{LM}(\vec{n})Y_{LM}^*(\vec{n}')$ by the relation:

$$\rho_L(r,r',\omega) = -\frac{1}{\pi}\mathrm{Im}\tilde{A}_L(r,r',\omega) . \qquad (4)$$

In accordance with the spectral expansion of the p-h propagator, the double transition density of Eq. (4) determines the energy-averaged strength function $S_L(\omega)$ related to an isoscalar non-spin-flip external field (probing operator) $V_{LM}(\vec{r}) = V_L(r)Y_{LM}(\vec{n})$ ($V_L(r)$ is supposed to be a real quantity):

$$S_L(\omega) = \int V_L(r)\,\rho_L(r,r',\omega)V_L(r')drdr' . \qquad (5)$$

It is noteworthy that due to the method of treating the spreading effect within PHDOM, the double transition density cannot be factorized in terms of one-body transition densities. Within the model, the strength function of Eq. (5) can be evaluated in a simpler way. For this aim, we define the so-called effective field $\tilde{V}_L(r,\omega)$ by the integral relation:

$$\int \tilde{A}_L(r,r',\omega)V_L(r')dr' = \int A_L(r,r',\omega)\tilde{V}_L(r',\omega)dr' . \tag{6}$$

In accordance with Eqs. (3) and (6), the effective field obeys the equation,

$$\tilde{V}_L(r,\omega) = V_L(r) + \frac{F(r)}{r^2}\int A_L(r,r',\omega)\tilde{V}_L(r',\omega)dr' , \tag{7}$$

which is simpler than Eq. (3). An alternative expression for the strength function follows from Eqs. (4)-(6):

$$S_L(\omega) = -\frac{1}{\pi} Im\, P_L(\omega) , \tag{8}$$

where $P_L(\omega)$ is the respective polarizability:

$$P_L(\omega) = \int V_L(r)A_L(r,r',\omega)\tilde{V}_L(r',\omega)drdr' . \tag{9}$$

Since the methods, used for describing hadron-nucleus scattering accompanied by GR excitation, employ only one-body transition density (see Refs. [11-13]), it is desirable to factorize approximately the double transition density of Eq. (4). It can be done in terms of the projected (one-body) transition density $\rho_{V_L}(r,\omega)$ related to a given probing operator and defined as follows [7]:

$$\rho_{V_L}(r,\omega) = \int \rho_L(r,r',\omega)V_L(r')dr'/S_L^{1/2}(\omega) . \tag{10}$$

Using Eqs. (5) to (10), one gets the following expressions, which are formally valid also in cRPA:

$$S_L(\omega) = \left(\int \rho_{V_L}(r,\omega)V_L(r)dr\right)^2 , \tag{11}$$

$$\frac{1}{r^2}\rho_{V_L}(r,\omega) = -\frac{1}{\pi}Im\tilde{V}_L(r,\omega)/(F(r)S_L^{1/2}(\omega)) . \tag{12}$$

All the considered quantities related to ISGMPRs are determined, in fact, by elements of the "free" p-h propagator $A_L = A_L^{n\bar{n}} + A_L^{p\bar{p}}$ (below the indexes "$n$" and "$p$" related to the neutron and proton subsystems are, as a rule, omitted for brevity sake). We note here that the basic Eqs. (3) and (7) are derived in the approximation: $(N - Z) \ll A$, where $A = N + Z$ is the number of nucleons. The expression for the "free" p-h propagator, in which the continuum and spreading effect are approximately taken into account, has been derived for closed-shell nuclei in a rather general form [3]. This expression adopted in Ref. [7] to describe within the PHDOM the Isoscalar Giant Monopole Resonance (i.e., to get the quantity $A_{L=0}(r,r',\omega)$) contains: the occupation numbers $n_\mu$; the single-particle radial bound-state wave functions $r^{-1}\chi_\mu(r)$ and energies $\varepsilon_\mu$, with $\mu = n_{r,\mu}, j_\mu, l_\mu, (\mu) \equiv j_\mu, l_\mu$ being the set of bound-state quantum numbers, and; the Green functions $g_{(\lambda)}(r,r',\varepsilon = \varepsilon_\mu \pm \omega)$ of the single-particle radial Schrodinger equation, which contains the complex term, $[-iW(\omega) + \mathcal{P}(\omega)]f_\mu f_{WS}(r))$, added to a mean field, with $W(\omega)$ and $\mathcal{P}(\omega)$ being the imaginary and real parts of the intensity of the energy-averaged p-h self-energy term responsible for the spreading effect, and $f_{WS}(r)$ and $f_\mu$ are the Woods-Saxon function and its diagonal matrix element, respectively. The expression for $A_L(r,r',\omega)$ can be obtained from that for $A_{L=0}(r,r',\omega)$, given in details in Ref. [7], by the substitution of the kinematic factors:

$$t^{L=0}_{(\lambda)(\mu)} = \frac{(2j_\mu+1)^{1/2}}{\sqrt{4\pi}}\delta_{(\lambda),(\mu)} \to t^L_{(\lambda)(\mu)} = \frac{1}{\sqrt{2L+1}}\langle(\lambda)\|Y_L\|(\mu)\rangle. \qquad (13)$$

The results of strength function calculations can be verified, using the weakly model-dependent energy-weighted sum rule $EWSR_L = \int \omega S_L(\omega)d\omega$ [1]:

$$EWSR_L = \frac{1}{4\pi}\frac{\hbar^2}{2M}A\langle\left(\frac{dV_L(r)}{dr}\right)^2 + L(L+1)\left(\frac{V_L(r)}{r}\right)^2\rangle. \qquad (14)$$

Here, the averaging $\langle...\rangle$ is performed over the nuclear density $n(r) = n^n(r) + n^p(r)$. In the next Section, the strength functions $S_L(\omega)$ calculated for the $^{208}$Pb nucleus are presented in terms of the relative energy-weighted strength functions (fractions of $EWSR_L$)

$$y_L(\omega) = \omega S_L(\omega)/EWSR_L, \qquad (15)$$

normalized by the condition $x_L = \int y_L(\omega)d\omega = 1$. (We omit the factor $(2L+1)$ in Eq. (14) in accordance with the definition used for strength functions of Eqs. (5) and (8).)

The choice of the radial part of probing operators $V_L(r)$ used for describing ISGMPRs within PHDOM depends on nature of the considered resonance. The position dependences of $V_{L=2,3}(r)$ are taken as $r^L$, because the Isoscalar Giant Quadrupole and Octupole Resonances (ISGQR and ISGOR, respectively) are related to main-tone collective excitations. The signature of these excitations is a nearly nodeless radial dependence of projected transition densities $\rho_{V_L}(r,\omega)$ (Eqs. (10) and (12)) taken at $\omega = \omega_{L(peak)}$ - the energy of the main maximum of the respective strength function. The Isoscalar Giant Monopole and Dipole Resonances (ISGMR and ISGDR, respectively) can be considered as the overtone excitations. The respective main tones are related to the spurious states (SS): the $0^+$ ground state and the $1^-$ state associated with center-of-mass motion. The signature of an arbitrary overtone is appearance of an extra-node in the radial dependence of the projected transition density taken at the main maximum of the overtone GR strength function. To suppress excitation of the above-mentioned spurious states, the radial dependence of the probing operators is taken as $V_{L=0}(r) = r^2 - \eta_0 \langle r^2 \rangle$ and $V_{L=1}(r) = r(r^2 - \eta_1 \langle r^2 \rangle)$. To avoid violation of PHDOM unitarity, the parameter $\eta_0$ is taken equal to unity [8]. Spurious isoscalar $1^-$ excitations are described by the polarizability $P_{L=1}^{SS}(\omega)$ related to the isoscalar dipole operator, having the radial part $V_{L=1}^{SS}(r) = r$. From the condition, that the mentioned polarizability related to this operator has a maximum at $\omega = \omega_{L=1}^{SS}$ close to zero excitation energy, one gets the strength of the isoscalar part of Landau-Migdal forces [17] (see Section III and Appendix A). Being determined by the effective field $\tilde{V}_{L=1}^{SS}(r, \omega \to \omega_{L=1}^{SS})$, the radial part of the $1^-$ spurious-state transition density $\rho_{L=1}^{SS}(r)/r^2$ (see Section III and Appendix A) might be used to find the parameter $\eta_1$ in the expression for the radial part of the second-order isoscalar dipole probing operator from the condition $\int V_{L=1}(r) \rho_{L=1}^{SS}(r) dr = 0$. Among the overtones of real isoscalar GRs, the overtones of monopole and quadrupole GRs (ISGMR2 and ISGQR2, respectively) have the lowest excitation energies [17, 7]. The radial part of the respective probing operators $V_L^{ov}(r) = r^2(r^2 - \eta_L^{ov} \langle r^2 \rangle)$ ($L = 0, 2$) contains the adopted parameter $\eta_L^{ov}$. To suppress main-tone excitation, this parameter can be found from the condition: $\int V_L^{ov}(r)\rho_{V_L}(r, \omega_{L(peak)})dr = 0$. The main-tone GR is placed at the distant low-energy "tail" of the corresponding overtone. All the above-considered overtones are related to "breathing" modes of nuclear excitations.

The ability to estimate quantitatively probabilities of direct one-nucleon decay of GRs belongs to unique features of PHDOM. Within the model (as in case of the non-standard cRPA version), the effective-field method is used for such estimations [6, 8, 9]. The strength function for direct one-nucleon decay of ISGMPR into the channel $\mu$, corresponding to population of the one-hole configuration $\mu^{-1}$ in the product nucleus, is determined by the squared amplitudes of "direct+semidirect" reactions induced by the external field $V_{LM}(\vec{r})$ [8]:

$$S_{L,\mu}^{\uparrow}(\omega) = \sum_{(\lambda)} n_\mu \, (t_{(\lambda)(\mu)}^L)^2 |\int \chi_{\varepsilon=\varepsilon_\mu+\omega,(\lambda)}^*(r)\tilde{V}_L(r,\omega)\chi_\mu(r)dr|^2 \,. \quad (16)$$

Here, $r^{-1}\chi_{\varepsilon>0,(\lambda)}(r)$ is the radial one-nucleon continuum-state wave function, having the standing-wave asymptotic behavior. Being normalized to the $\delta$-function of the energy in the $W = \mathcal{P} = 0$ limit, this wave function obeys the mentioned Schrodinger equation, in which the above-described complex term is added to the mean field. The partial branching ratio for direct one-nucleon decay of the ISGMPR into the channel $\mu$, $b_{L,\mu}^{\uparrow}$, is determined by the strength functions of Eqs. (16), (8), and (9):

$$b_{L,\mu}^{\uparrow} = \int_{\omega_{12}} S_{L,\mu}^{\uparrow}(\omega)d\omega \,/\, \int_{\omega_{12}} S_L(\omega)d\omega \,. \quad (17)$$

Here, $\omega_{12} = \omega_1 - \omega_2$ is an energy interval, that includes the considered GR. The total branching ratio, $b_{L,tot}^{\uparrow} = \sum_\mu b_{L,\mu}^{\uparrow}$ (summation on the neutron and proton subsystems is also included), determines the branching ratio for statistical (mainly neutron) decay: $b_L^\downarrow = 1 - b_{L,tot}^\uparrow$. Note that in the cRPA limit ($W = \mathcal{P} = 0$) $b_{L,tot}^\uparrow = 1$ and $b_L^\downarrow = 0$.

### III. DESCRIPTION OF ISGMPRs

Within the current PHDOM version employed for describing the main characteristics of ISGMPRs in the medium-heavy closed-shell nuclei, the following input quantities are used: 1) a realistic (Woods-Saxon-type) phenomenological partially self-consistent mean field $U(x)$ (described in details in Ref. [18]); 2) the non-spin-flip part of Landau-Migdal p-h interaction (Eq. (2)) with the isovector $F'$ and isoscalar $F(r)$ strengths related to the mean field due to approximate restoration of isospin symmetry and translation invariance of the model Hamiltonian, respectively, and; 3) the phenomenological imaginary part $W(\omega)$ of the intensity of an energy-averaged p-h self-energy term responsible for the spreading effect.

1) The mean field $U(x)$ contains the isoscalar part $U_0(x)$, including the central and spin-orbit terms, the isovector and Coulomb parts $U_1(x)$ and $U_C(x)$, respectively, are taken as:

$$U(x) = U_0(x) + U_1(x) + U_C(x), \qquad (18)$$

$$U_0(x) = -U_0 f_{WS}(r, R, a) + U_{ls}\frac{1}{r}\frac{df_{WS}}{dr}\vec{l}\vec{s}, \qquad (19)$$

$$U_1(x) = \frac{1}{2}\tau^{(3)}v(r), \quad U_C(x) = \frac{1}{2}(1 - \tau^{(3)})U_C(r). \qquad (20)$$

Here, $R = r_0 A^{1/3}$, $r_0$ and $a$ are the size and diffuseness parameters, respectively; $U_0$ and $U_{ls}$ are the strength parameters related to the isoscalar central and spin-orbit terms, respectively (the quantity $\vec{l}\vec{s}$ is taken in the fraction of $\hbar^2$); $v(r) = 2F'n^{(-)}(r)$ is the symmetry potential calculated self-consistently via the neutron-excess density $n^{(-)}(r) = n^n(r) - n^p(r)$, and; $U_C(r)$ is the mean Coulomb field which is also calculated self-consistently via the proton density $n^p(r)$.

2) The isoscalar and isovector strengths of the non-spin-flip part of Landau-Migdal p-h interaction are taken as $F(r) = Cf(r)$ and $F' = Cf'$, $C = 300\, MeVfm^3$. From the above-given expression for the symmetry potential, it follows that Landau-Migdal parameter $f'$ can be related to mean-field parameters. The isoscalar strength $f(r)$ is parameterized in accordance with Ref. [19] as:

$$f(r) = f^{ex} + (f^{in} - f^{ex})f_{WS}(r). \qquad (21)$$

The small parameter $f^{in}$ is usually taken as an universal quantity, while the main parameter $f^{ex}$ in Eq. (21) is found for each considered nucleus from the condition, that the energy $\omega_{L=1}^{SS}$ of the spurious isoscalar dipole state is close to zero (see Section II). The spurious-state energy and strength can be found by parameterization of the inverse polarizability of Eq. (9) related to spurious isoscalar dipole excitations:

$$[P_{L=1}^{SS}(\omega \to 0)]^{-1} = [\omega^2 - (\omega_{L=1}^{SS})^2]/(2x_{L=1}^{SS} EWSR_{L=1}^{SS}). \qquad (22)$$

In this expression, which follows from the spectral expansion for the p-h Green function taken at low excitation energies (see Eqs. (A1), and (A2) of Appendix A), the quantity $x_{L=1}^{SS}$ is the spurious-state fraction of the respective energy-weighted sum rule of Eq. (14), $x_{L=1}^{SS} = \omega_{L=1}^{SS}(M_{L=1}^{SS})^2/EWSR_{L=1}^{SS}$, with $M_{L=1}^{SS}$ being the spurious-state excitation amplitude. The reasonable choice of parameters $f^{ex}$ and $f^{in}$ means that within the model used, the spurious-state energy is close to zero and the spurious-state fraction of $EWSR_{L=1}^{SS}$ is close to unity. Here, we show also the expression for the $1^-$ spurious-state transition density (see Eqs. (A2), and (A3) of Appendix A)

$$\rho_{L=1}^{SS}(r) = r^2 \left(\widetilde{V}_{L=1}^{SS}(r, \omega \to 0) - V_{L=1}^{SS}(r)\right) M_{L=1}^{SS} / \left(F(r)P_{L=1}^{SS}(\omega \to 0)\right), \qquad (23)$$

previously used (Section II).

Actually, the outlined method for the $1^-$ spurious state description can be used for the evaluation, within cRPA, of the energy $\omega_L^{coll}$, strength

$x_L^{coll} EWSR_L/\omega_L^{coll}$, and transition density $\rho_L^{coll}(r)$ of any isoscalar collective state below the nucleon-escape threshold. In this case, the calculated inverse polarizability can be presented in the form

$$[P_L(\omega \to \omega_L^{coll})]^{-1} = (\omega - \omega_L^{coll})\omega_L^{coll}/(x_L^{coll} EWSR_L), \qquad (24)$$

which allows to find the collective-state energy and strength. The latter is usually described in terms of the reduced $EL$-transition probability $(B(EL), 0^+ \to L^\pi)/e^2$, where $e$ is the proton charge [1]. This quantity is related to the collective-state strength as follows:

$$x_L^{coll} EWSR_L/\omega_L^{coll} = \frac{4}{2L+1}(B(EL), 0^+ \to L^\pi)/e^2. \qquad (25)$$

Factor 4 in this expression appears due to isobaric structure of the external field related to $EL$-transitions ($L \neq 1$): $V_{EL} \sim \frac{e}{2}(1 - \tau^{(3)})$. Eq. (23) taken at $\omega \to \omega_L^{coll}$ can be directly used to get the expression for the collective-state transition density in terms of the related effective field.

3) Following Refs. [6-9], we take the imaginary part of the intensity of the energy-averaged p-h self-energy term responsible for the spreading effect as three-parametric function of the excitation energy:

$$2W(\omega) = \begin{cases} 0, & \omega < \Delta; \\ \alpha(\omega - \Delta)^2/[1 + (\omega - \Delta)^2/B^2], & \omega \geq \Delta. \end{cases} \qquad (26)$$

Here the adjustable parameters $\alpha$, $\Delta$ and $B$ can be called as the strength, gap and saturation parameters, respectively. The use of Eq. (26) for evaluation of the real part, $\mathcal{P}(\omega)$, by means of the proper dispersive relationship [3] leads to a rather cumbersome expression, which can be found in Ref. [20]. The above-mentioned adjustable parameters are found from the PHDOM-based description of observable total width (FWHM) and, to some extent, peak energy for considered ISGMPRs in a given nucleus.

The strength ($U_0$, $U_{ls}$, $f'$) and geometrical ($r_0$, $a$) mean-field parameters together with the parameters $f^{ex}$, $f^{in}$ of Eq. (21) are the independent input data used in implementation of PHDOM for describing ISGMPRs in medium-heavy closed-shell nuclei. For the $^{208}$Pb nucleus taken as an appropriate example, the above-listed parameters are found from a description of observable single-quasiparticle spectra in the respective even-odd and odd-even nuclei. Table 1 contains the mean-field parameters, the p-h interaction parameters and also adjustable ("spreading") parameters. The latter determine the quantities $W(\omega)$ and $\mathcal{P}(\omega)$ (Fig. 1). Thus, all the main characteristics of ISGMPRs (including $L = 0, 2$ overtones) are described without the use of additional parameters.

In presenting the ISGMPR main characteristics calculated within the current PHDOM version for the $^{208}$Pb nucleus, we start from the relative energy-weighted strength functions $y_L(\omega)$ of Eq. (15). These functions are shown in Fig. 2 for ISGMR and ISGMR2, Fig. 3 for ISGQR and ISGQR2, and Fig. 4 for ISGDR and ISGOR. In Tables 2, 3, the following ISGMPRs parameters deduced from calculated strength functions $S_L(\omega)$ of Eqs. (8), and (9) are given together with available experimental data: the fraction of $EWSR_L$, $x_L$, evaluated for a large excitation-energy interval $\omega_{12}$; the main peak energy (energies), $\omega_{L(peak)}$; the centroid energy $\overline{\omega}_L$ evaluated for a given energy interval within PHDOM and cRPA in a comparison with the results of the self-consistent microscopic approach of HF-based RPA using the SkT1 Skyrme force associated with m$^*$/m $= 1$ [10]; the total width (the full width at half maximum) $\Gamma_{L(FWHM)}$, and; the parameters $\eta_L$ and $\eta_L^{ov}$ used in the definition of the respective probing operator. The ISGMPRs parameters evaluated within cRPA (Tables 2, 3) are obtained with the use of the small (energy-independent) "technical" value $2W = 10$ keV.

The $1^-$ spurious-state parameters deduced from the calculated polarizability $P_{L=1}^{SS}(\omega)$ of Eqs. (9), (22) $\omega_{L=1}^{SS} \cong 20$ keV and $x_{L=1}^{SS} \cong 93$ % are related to the chosen parameters $f^{ex}$ and $f^{in}$. As an example of describing within cRPA low-energy isoscalar collective states (phonons) according to Eqs. (24), and (25), we present in Table 4 the calculated characteristics of $3^-$ and $2^+$ collective states in $^{208}$Pb.

The next main characteristic of the considered GRs is the projected transition density of Eq. (10), $\rho_{V_L}(r,\omega)$. Evaluated within PHDOM these densities taken at the peak energy of respective ISGMPR in $^{208}$Pb are shown in Figs. 5-8. In Figs. 7 and 8, the $1^-$ spurious-state and low energy $3^-$ transition densities calculated within cRPA are also shown.

Turning to direct one-nucleon decay of the considered ISGMPRs (Section II), we show the partial and total branching ratios, $b^{\uparrow}_{L,\mu}$ and $b^{\uparrow}_L$, evaluated within the PHDOM for $^{208}$Pb. The neutron branching ratios are given in Table 5 with indication of the respective excitation-energy intervals. The calculated partial branching ratios for direct one-proton decay of ISGDR are compared with available experimental data in Table 6.

## IV. DISCUSSION OF RESULTS. CONCLUSIVE REMARKS

In the previous Section, we presented the PHDOM-based description of main properties of six ISGMPRs in the $^{208}$Pb nucleus. The cRPA-based description of a few low-energy bound isoscalar (p-h)-type states (including the $1^-$ spurious state related to centre-of-mass motion) is also given. Most of input quantities (enough for a cRPA-based description of isoscalar and isovector non-spin-flip GRs), namely, the mean-field parameters and (p-h)-interaction strengths are taken from independent data with accounting for fundamental symmetries of the model Hamiltonian (Table 1). Only the specific ("spreading") parameters have been adjusted to get within the model a reasonable description of experimental total width and, to some extent, the peak energy of considered GRs (Table 2). The strength-function calculations are verified by the use of respective energy-weighted sum rules (Tables 2-4). The spreading shift related to the centroid energy of considered GRs is found relatively small (0.2-0.6 MeV), as it follows from comparing the centroid energies evaluated within the PHDOM and the cRPA (Table 2). The spreading shift related to the ISGMPR peak-energy (Table 2) is in a qualitative agreement with the energy dependence of the real (dispersive) part of the intensity of the energy-averaged (p-h) self-energy term $\mathcal{P}(\omega)$ (Fig. 1). It is noteworthy that the observed area of the ISGQR2 strength concentration (around 26 MeV [22]) is in agreement with the respective data of Table 2. We note also that the centroid energies of ISGMPRs calculated within cRPA and the microscopic approach of HF-based RPA using the SkT1 Skyrme force associated with $m^*/m = 1$ [10] are close (Table 2).

The evaluated ISGMPR parameters given in Tables 2, 3 are deduced from strength functions $S_L(\omega)$ calculated in a large excitation energy interval. It is convenient to compare the strength distributions for various GRs in terms of the relative energy-weighted strength functions $y_L(\omega)$ (Figs. 2-4). As expected, the degree of strength concentration is decreased with increasing GR excitation energy.

Main characteristics of ISGMPR include the projected transition density $\rho_{V_L}(r, \omega)$ considered in a large excitation energy interval. The radial dependence of $\rho_{V_L}(r, \omega = \omega_{L(peak)})$ (Figs. 5-8) can be considered as a signature of considered GR. The nodeless radial dependence is related to main-tone GRs (ISGQR and ISGOR); the one-node radial dependence is related to first-order-overtone GRs (ISGMR, ISGDR, ISGQR2); the two-node radial dependence is related to second-order-overtone GRs (ISGMR2). The radial dependence of the $1^-$ spurious-state transition density $\rho_{L=1}^{SS}(r)$ (Fig. 7) exhibits, naturally, nodeless radial dependence.

A possibility to estimate the branching ratios for direct one-nucleon decay of an arbitrary GR related to unique features of the PHDOM-based approach is shown. The respective relations given in Section II are, actually, obtained under assumption of a purely single-hole nature of the product-nucleus states that are populated in the decay process. Therefore, the calculated partial branching ratios can be considered as an upper limit of possible values. In Table 5, the evaluated branching ratios for direct one-neutron decay of four ISGMPRs in the $^{208}$Pb nucleus are given together with available experimental data. An approximately two-fold excess of the calculated values above the respective experimental values for ISGMR and ISGDR is worth noting. This note is also valid for the calculated values of the branching ratios for direct one-proton decay of ISGDR (Table 6). However, the description of experimental data is markedly improved upon multiplying the calculated branching ratios $b_{L=1,\mu}^{\uparrow}$ by the experimental values of spectroscopic factors $S_\mu$ for proton-hole states of the product nucleus $^{207}$Tl. (The experimental spectroscopic factors $S_\mu$ are close to unity for the majority of neutron-hole states of the $^{207}$Pb nucleus, which are indicated in Table 5).

In conclusion, the particle-hole dispersive optical model, was implemented for describing main properties of Isoscalar Giant Multipole Resonances up to $L = 3$ in medium-heavy closed-shell nuclei. The overtones of the monopole and quadrupole isoscalar giant resonances were also studied. The main properties, considered in a large excitation-energy interval, include the following energy-averaged quantities: (i) the strength function related to an appropriate probing operator; (ii) the projected one-body transition density (related to the corresponding operator), and; (iii) partial probabilities of direct one-nucleon decay. Unique abilities of PHDOM where conditioned by a joint description of the main relaxation processes of high-energy p-h configurations associated with a given giant resonance. Two processes, Landau damping and coupling the mentioned configurations to the single-particle continuum, were described microscopically in terms of Landau-Migdal p-h interaction and a phenomenological mean field, partially consistent with this interaction. Another mode, the coupling to many quasiparticle states (the spreading effect) was described phenomenologically in terms of the imaginary part of the properly parameterized energy-averaged p-h self-energy term. The imaginary part determines the real one via a microscopically-based dispersive relationship.

The model parameters related to a mean field and p-h interaction were taken from independent data accounting for the isospin symmetry and translation invariance of the model Hamiltonian. Parameters of the imaginary part of the strength of self-energy term were adjusted to reproduce in PHDOM-based calculations of ISGMPR total widths for the considered closed-shell nucleus. The calculation results obtained for the $^{208}$Pb nucleus, taken as an example, were compared with available experimental data. Some of the results were compared with those obtained in microscopic HF-based RPA calculations using Skyrme force. These comparisons indicate that PHDOM is a powerful tool for describing ISGMPRs in medium-heavy closed-shell nuclei. The implementation of the PHDOM-based approach to describing ISGMPRs in other medium-heavy nuclei is in progress.

## AKNOWLEDGEMENTS

This work is partially supported by the Russian Foundation for Basic Research, under Grant no. 19-02-00660 (M.L.G., B.A.T., M.H.U.), by the US Department of Energy, under Grant no. DE-FG03-93ER40773 (S.S.), and by the Competitiveness Program for National Research Nuclear University-MEPhI (M.H.U.).

## APPENDIX A

### *A cRPA description of isoscalar bound (p-h)-type states*

**1.** The description is based on the spectral expansion of the cRPA p-h Green function, whose radial components, $\tilde{A}_L^{cRPA}(r,r',\omega)$, satisfying Eq. (1) in neglecting the spreading effect, can be presented in the form:

$$\tilde{A}_L^{cRPA}(r,r',\omega) = \sum_c \rho_{c,L}(r)\rho_{c,L}(r') \left[\frac{1}{\omega-\omega_{c,L}+i0} - \frac{1}{\omega+\omega_{c,L}-i0}\right]. \quad (A1)$$

Here, $\omega_{c,L}$ and $\rho_{c,L}(r)$ are, respectively, the energy and radial (one-dimensional) transition density of (p-h)-type states. These are (latter) normalized to unity for bound states and to the δ-function of the energy for continuum states.

**2.** In case of the spurious isoscalar $1^-$ state, having within the model the energy $\omega_{L=1}^{SS}$ close to zero, the expression for the polarizability of Eqs. (6) and (9) related to the external field $V_{L=1}^{SS}(r) = r$ and taken in the cRPA limit follows from (A1):

$$P_{L=1}^{SS}(\omega \to 0) = 2\omega_{L=1}^{SS}(M_{L=1}^{SS})^2 / \left(\omega^2 - \left(\omega_{L=1}^{SS}\right)^2\right). \quad (A2)$$

Here, $M_{L=1}^{SS} = \int \rho_{L=1}^{SS}(r) r dr$ is the spurious-state excitation amplitude. The use of the inverse polarizability of Eq. (22) is more convenient for searching the spurious-state energy and fraction of $EWSR_{L=1}^{SS}$.

**3.** The expression for the spurious-state transition density $\rho_{L=1}^{SS}(r)$ of Eq. (23) follows from Eqs. (6) and (7) (considered in the cRPA limit) and Eqs. (A1) and (A2) taken at $\omega \to \omega_{L=1}^{SS} \to 0$:

$$\tilde{V}_{L=1}^{SS}(r, \omega \to 0) - V_{L=1}^{SS}(r) = \frac{F(r)}{r^2} \rho_{L=1}^{SS}(r) M_{L=1}^{SS} 2\omega_{L=1}^{SS} / \left(\omega^2 - \left(\omega_{L=1}^{SS}\right)^2\right). \quad (A3)$$

The radial dependence of $\rho_{L=1}^{SS}(r)/r^2$ is expected to be close to the "ideal" spurious-state transition density, which is proportional to the radial gradient of nuclear density. As a result, the parameter $\eta_1$ in the expression for the probing operator $V_{L=1}(r)$ (Section II) is expected to be close to the widely-used quantity 5/3.

**4.** The main characteristics of isoscalar collective bound states, $\omega_L^{coll}$, $M_L^{coll}$ (or $x_L^{coll}$), and $\rho_L^{coll}(r)$ can be obtained from Eqs. (A2) and (A3), related to the respective external field $V_L(r)$ and taken at $\omega \to \omega_L^{coll}$. In such a case, the ratio $2\omega_L^{coll}/\left(\omega^2 - \left(\omega_L^{coll}\right)^2\right)^2$ is going to $1/(\omega - \omega_L^{coll})$ (see, e.g., Eq. (24)).

**Table 1.** The list of mean-field and adjustable model parameters (notations are given in the text) used in calculations for $^{208}$Pb.

| $U_0$, MeV | $U_{ls}$, MeV fm$^2$ | $f'$ | $r_0$, fm | $a$, fm | $f^{ex}$ | $f^{in}$ | $\alpha$, MeV$^{-1}$ | B, MeV | $\Delta$, MeV |
|---|---|---|---|---|---|---|---|---|---|
| 55.74 | 33.35 | 0.976 | 1.21 | 0.63 | -2.66 | 0.0875 | 0.20 | 4.5 | 3.0 |

**Table 2.** The ISGMPRs parameters calculated for $^{208}$Pb together with available experimental data (notations are given in the text).

| $L, \eta_L$ | $\omega_1 - \omega_2$, MeV | $x_L$, % | $\bar{\omega}_L$, MeV | $\omega_{L(peak)}$, MeV | $\Gamma_{L(FWHM)}$, MeV | |
|---|---|---|---|---|---|---|
| 0<br>$\eta_0 = 1$ | 5 – 35 | 100 | 13.9 | 14.2 | 1.3 | cRPA |
| | 5 – 35 | 103 | 14.5 | 14.2 | 4.2 | PHDOM |
| | 8 – 20 | 99±15 | - | 13.96±0.20<br>13.6±0.2 | 2.88±0.20<br>3.6±0.4 | Expt. [14]<br>Expt. [21] |
| | 7 – 60 | - | 13.92 | - | - | SkT1 |
| 1<br>(LE)<br>$\eta_1 = 1.72$ | 5 – 15 | 18 | 9.5 | - | - | cRPA |
| | 5 – 15 | 18 | 9.7 | 6.3 ; 11.8 | - | PHDOM |
| | - | 24±15 | - | 13.26±0.30 | 5.68±0.50 | Expt. [14] |
| 1<br>(HE)<br>$\eta_1 = 1.72$ | 15 – 35 | 81 | 22.9 | - | - | cRPA |
| | 15 – 35 | 83 | 23.5 | 23.9 | 7.0 | PHDOM |
| | 8 – 35 | 88±15 | - | 22.20±0.30<br>22.5±0.3<br>22.1±0.3 | 9.39±0.35<br>10.9±0.9<br>3.8±0.8 | Expt. [14]<br>Expt. [21]<br>Expt. [19] |
| | 16 – 60 | - | 23.40 | - | - | SkT1 |
| 2 | 5 – 35 | 86 | 11.0 | 10.7 | 0.1 | cRPA |
| | 5 – 35 | 90 | 11.3 | 10.6 | 2.7 | PHDOM |
| | 8 – 35 | 100±13 | - | 10.89±0.30<br>10.9±0.3 | 3.0±0.3<br>3.1±0.3 | Expt. [14]<br>Expt. [23] |
| | 7 – 60 | - | 10.55 | - | - | SkT1 |
| 3<br>(LE) | 5 – 15 | 17 | 8.4 | - | - | cRPA |
| | 5 – 15 | 19 | 8.7 | 5.8 | 1.1 | PHDOM |
| 3<br>(HE) | 15 – 35 | 60 | 19.8 | 18.9 | 0.3 | cRPA |
| | 15 – 35 | 61 | 20.7 | 19.5 | 3.7 | PHDOM |
| | 8 – 35 | 70±14 | - | 19.6±0.5<br>19.1±1.1 | 7.4±0.6<br>5.3±0.8 | Expt. [14]<br>Expt. [24] |
| | 15 – 60 | - | 19.34 | - | - | SkT1 |

**Table 3.** The overtone parameters calculated for ISGMR2 and ISGQR2 in $^{208}$Pb (see text).

| $L, \eta_L$ | $\omega_1 - \omega_2$, MeV | $x_L$, % | $\bar{\omega}_L$, MeV | $\omega_{L(peak)}$, MeV | |
|---|---|---|---|---|---|
| 0<br>$\eta_0^{ov} = 2.43$ | 5 – 15 | 17 | 10.9 | - | cRPA |
| | 5 – 15 | 18 | 11.1 | 10.0 | PHDOM |
| | 15 – 45 | 81 | 27.1 | - | cRPA |
| | 15 – 45 | 81 | 27.6 | 33.6 | PHDOM |
| 2<br>$\eta_2^{ov} = 1.73$ | 5 – 15 | 22 | 11.7 | - | cRPA |
| | 5 – 15 | 22 | 11.4 | 10.7 | PHDOM |
| | 15 – 45 | 78 | 26.0 | - | cRPA |
| | 15 – 45 | 81 | 26.5 | 32.1 | PHDOM |

**Table 4.** Evaluated within cRPA parameters of isoscalar collective low-energy $3^-$ and $2^+$ states in $^{208}$Pb in comparison with respective experimental data [25] (see text).

| $J^\pi$ | $\omega_L^{coll}$, MeV | $\omega_{L,expt}^{coll}$, MeV | $x_L^{coll}$, % | $x_{L,expt}^{coll}$, % |
|---|---|---|---|---|
| $3^-$ | 2.40 | 2.61 | 21 | 11 |
| $2^+$ | 4.42 | 4.09 | 14 | - |

**Table 5.** The partial and total branching ratios for direct one-neutron decay of the ISGMPR into the channel $\mu$. The evaluated within PHDOM branching ratios (in %) for $^{208}$Pb are given with indication of the respective excitation-energy intervals $\omega_{12}$ (in MeV) (see text).

| | $b^{\uparrow}_{L=0,\mu}$ | $b^{\uparrow}_{L=1,\mu}$ | $b^{\uparrow}_{L=2,\mu}$ | $b^{\uparrow}_{L=3,\mu}$ |
|---|---|---|---|---|
| $\mu^{-1} \setminus \omega_{12}$ | 12.5 – 15.5 [26] | 20 – 25 [27] | 9 – 12 | 16 – 23 |
| $3p_{1/2}$ | 3.6 | 1.1 | 2.8 | 1.6 |
| $2f_{5/2}$ | 18.0 | 5.4 | 1.5 | 5.9 |
| $3p_{3/2}$ | 7.5 | 2.6 | 5.8 | 3.8 |
| $1i_{13/2}$ | 0.8 | 11.4 | 0.2 | 5.9 |
| $2f_{7/2}$ | 26.6 | 9.3 | 0.2 | 13.3 |
| $\sum_{\mu} b^{\uparrow}_{L,\mu}$ | 56.5 | 29.8 | 10.5 | 30.5 |
| $\left(\sum_{\mu} b^{\uparrow}_{L,\mu}\right)_{expt}$ | 22 ± 6 [26]  14.3 ± 3 [28] | 23 ± 5 [27]  10.5 [22] | - | - |
| $b^{\uparrow,n}_{L}$ | 56.7 | 66.8 | 10.6 | 37.5 |

**Table 6.** The branching ratios (in %) for direct one-proton decay of the ISGDR in $^{208}$Pb evaluated within PHDOM for the excitation-energy intervals $\omega_{12} = 20 - 25$ MeV (see text).

| $\mu^{-1}$ | $b^{\uparrow}_{L=1,\mu}$ | $S_\mu$ [29] | $S_\mu \cdot b^{\uparrow}_{L=1,\mu}$ | $\left(b^{\uparrow}_{L=1,\mu}\right)_{expt}$ [27] |
|---|---|---|---|---|
| $3s_{1/2}$ | 3.4 | 0.55 | 1.9 | 2.3 ± 1.1 |
| $2d_{3/2}$ | 3.0 | 0.57 | 1.7 | |
| $1h_{11/2}$ | 0.2 | 0.58 | 0.1 | 1.2 ± 0.7 |
| $2d_{5/2}$ | 4.1 | 0.54 | 2.2 | |
| $\sum_\mu b^{\uparrow}_{L,\mu}$ | 10.7 | - | 5.9 | 3.5 ± 1.8 |

**Fig. 1.** The phenomenological imaginary part $W(\omega)$ (solid line) and real part $\mathcal{P}(\omega)$ (thin line) of the energy-averaged p-h self-energy term intensity, evaluated for $^{208}$Pb (the "spreading" parameters used are given in Table 1).

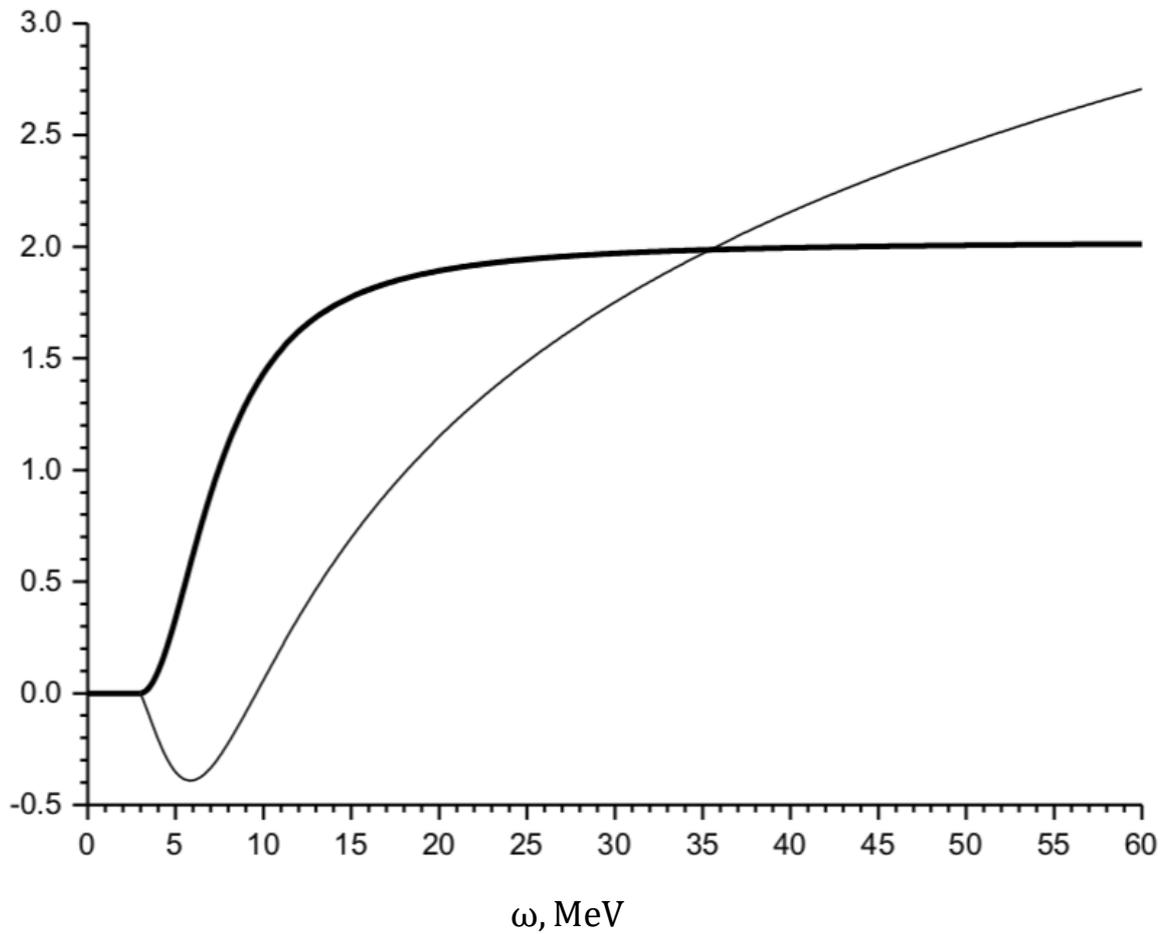

**Fig. 2.** The relative energy-weighted strength functions calculated within PHDOM for ISGMR (solid line) and ISGMR2 (thin line) in $^{208}$Pb.

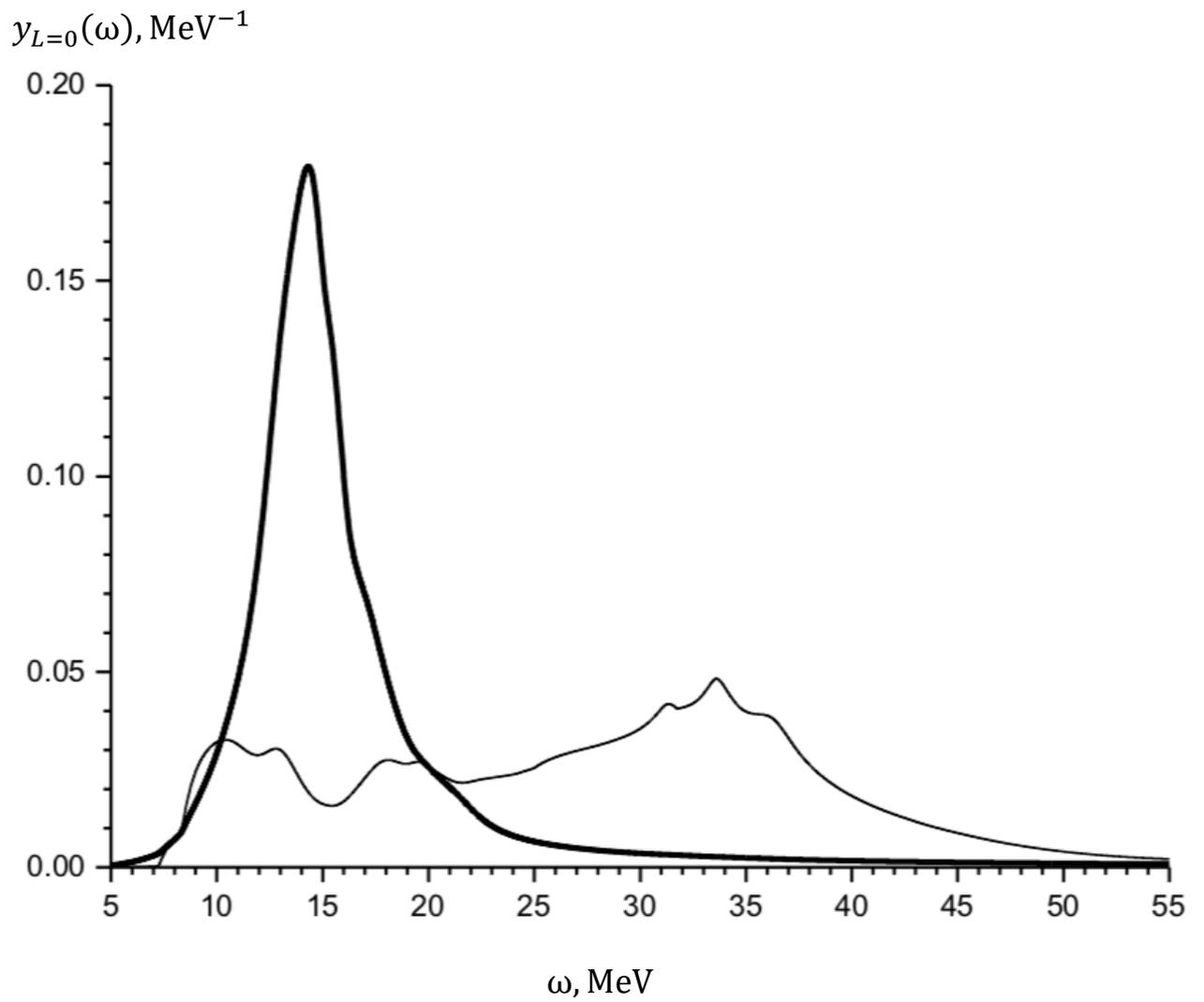

**Fig. 3.** The same as in Fig.2, but for ISGQR and ISGQR2.

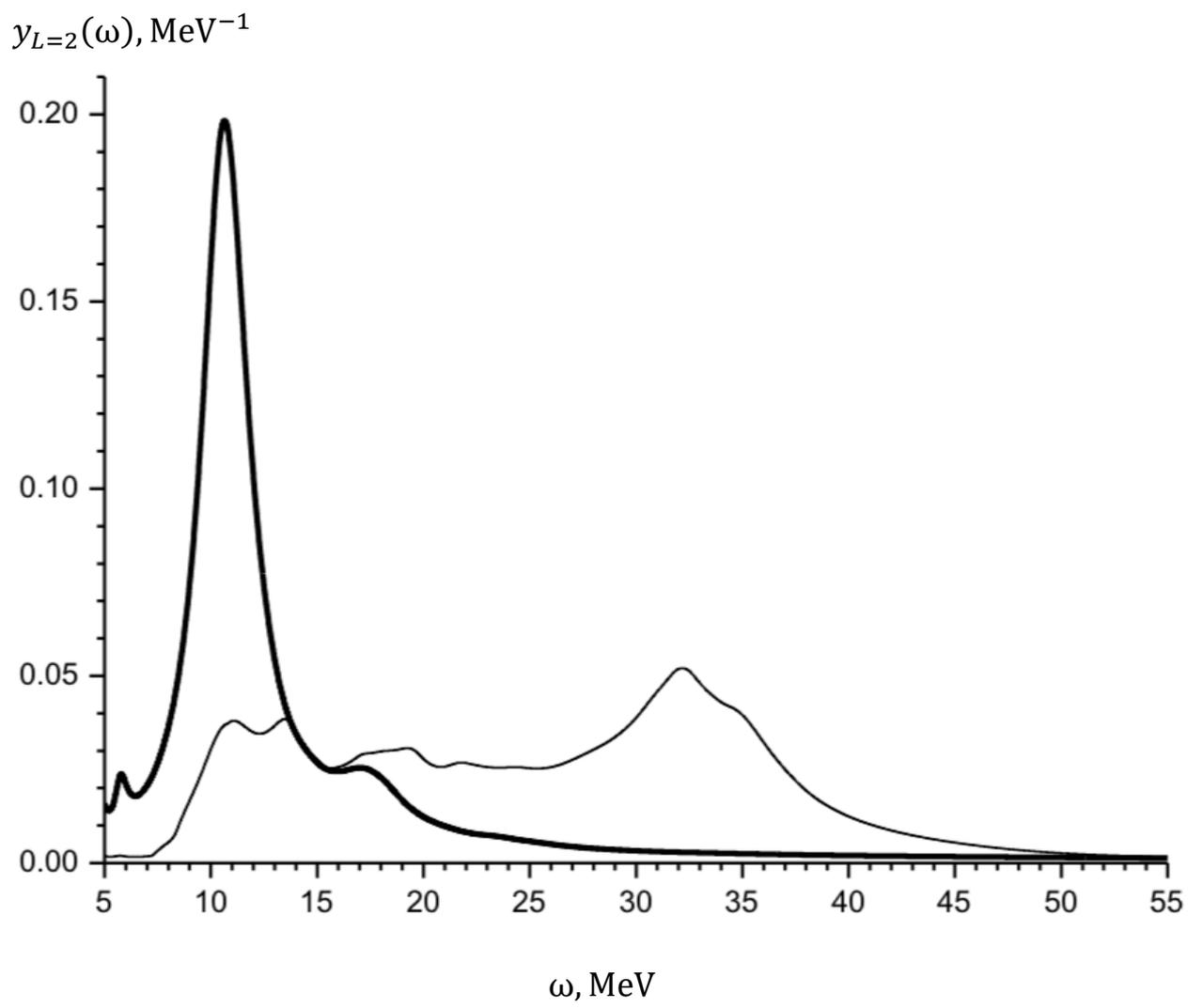

**Fig. 4.** The same as in Fig.2, but for ISGDR (solid line) and ISGOR (thin line).

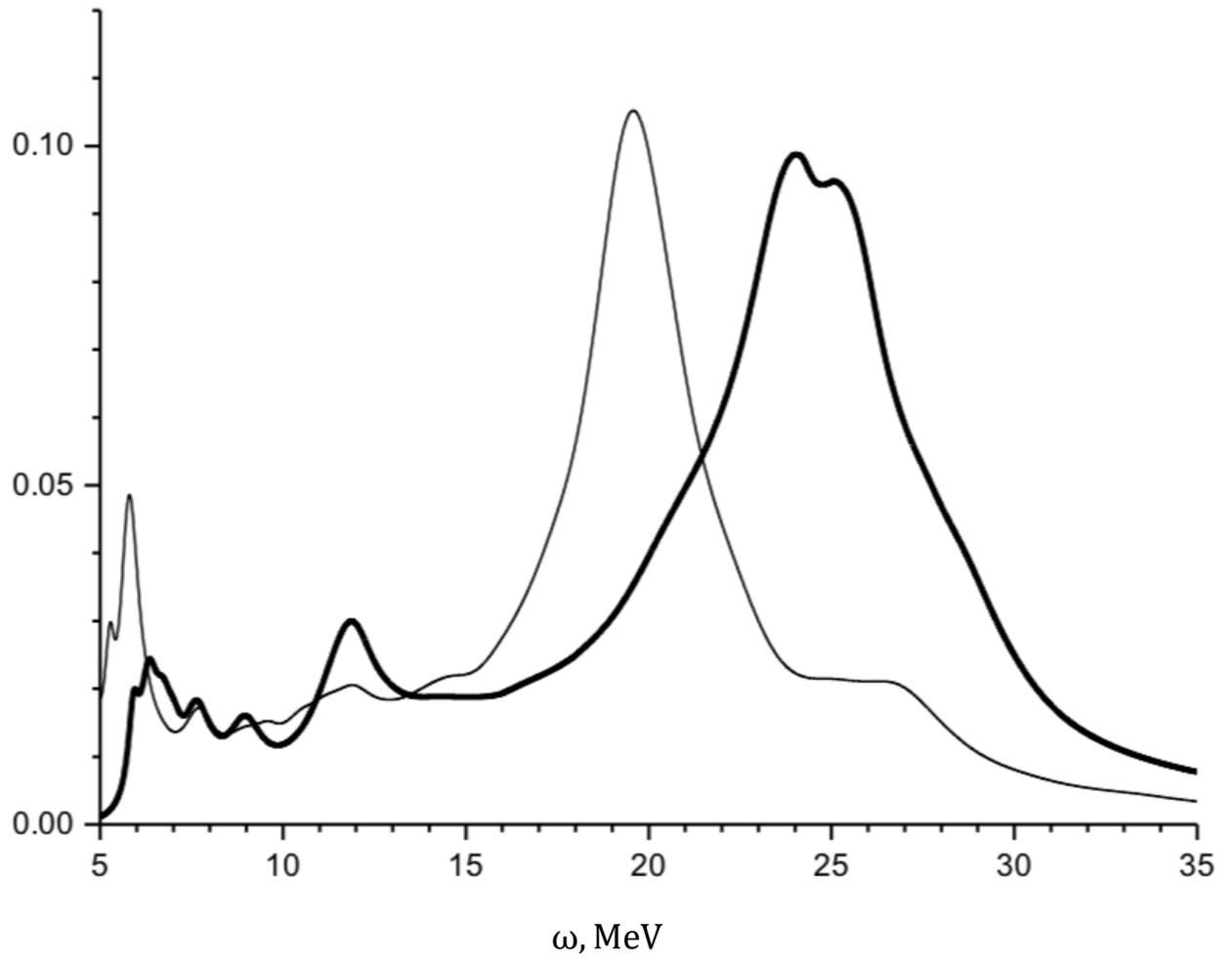

**Fig. 5.** The projected transition densities taken at the resonance peak-energy and calculated within PHDOM for ISGMR (solid line) and High-Energy ISGMR2 (thin line) in $^{208}$Pb.

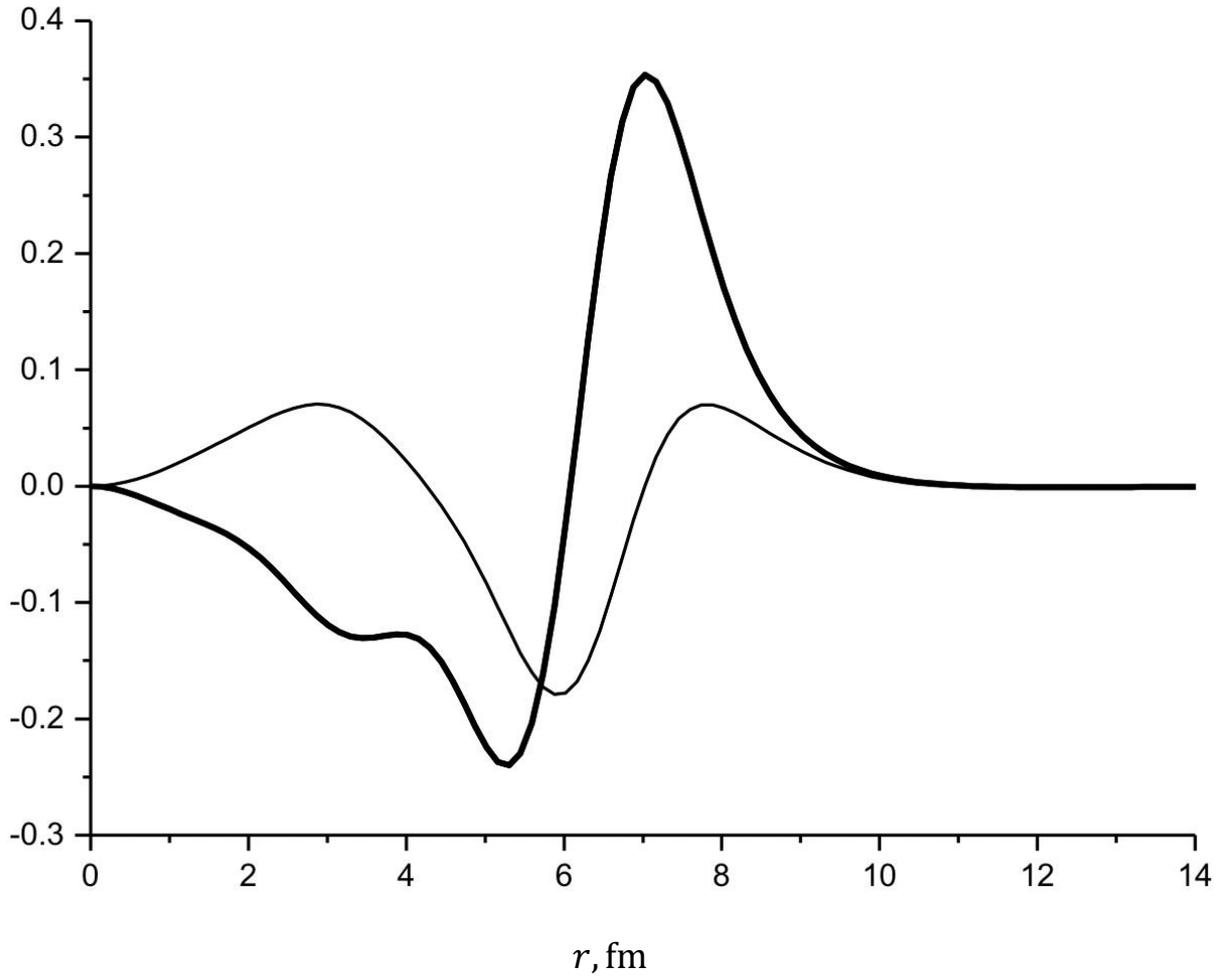

$\rho_{V_{L=0}}(r, \omega_{L=0(peak)})$, fm$^{-1}$MeV$^{-1/2}$

$r$, fm

**Fig. 6.** The same as in Fig.5, but for ISGQR and High-Energy ISGQR2.

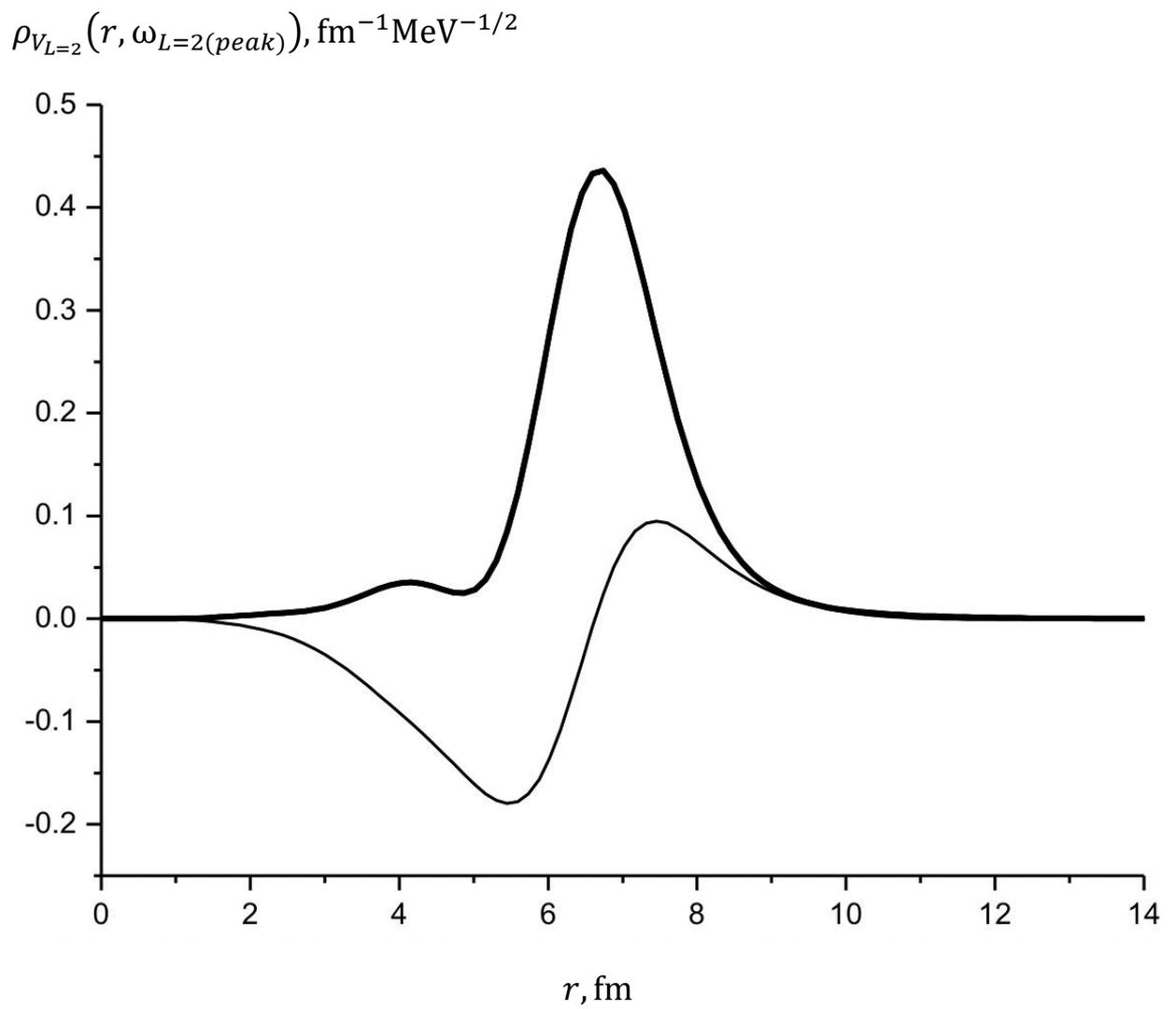

**Fig. 7.** The projected transition density taken at the resonance peak-energy and calculated within PHDOM for High-Energy ISGDR (solid line), and the $1^-$ spurious-state transition density calculated within cRPA (thin line) for $^{208}$Pb.

$\rho_{V_{L=1}}(r, \omega_{L=1(peak)})$, fm$^{-1}$MeV$^{-1/2}$

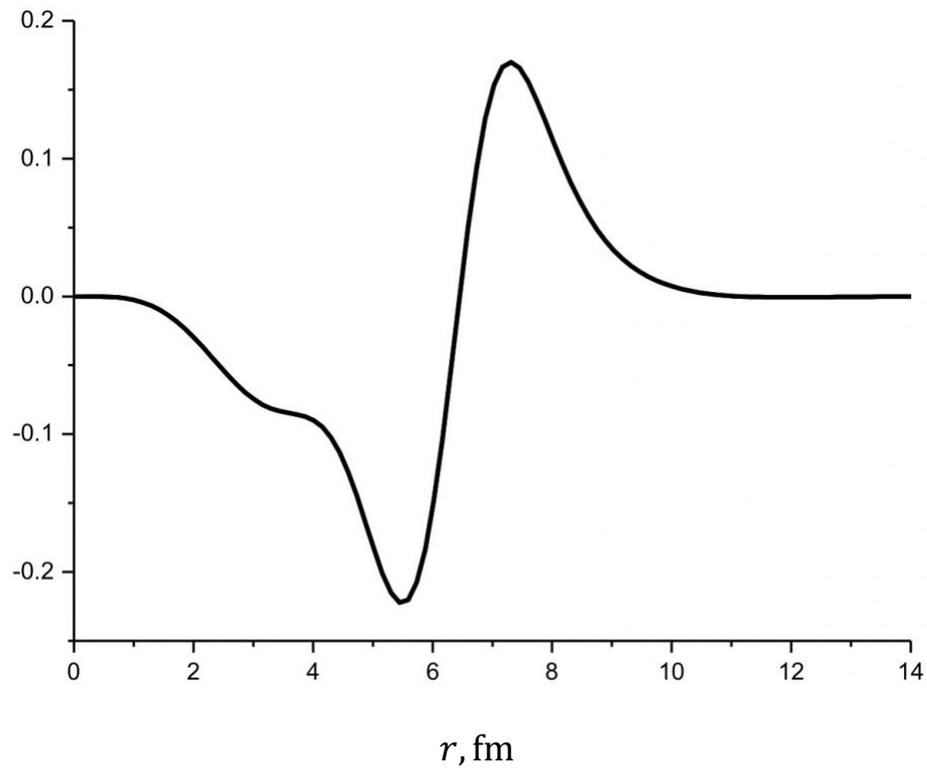

$r$, fm

$\rho^{SS}_{L=1}(r)$, fm$^{-1}$

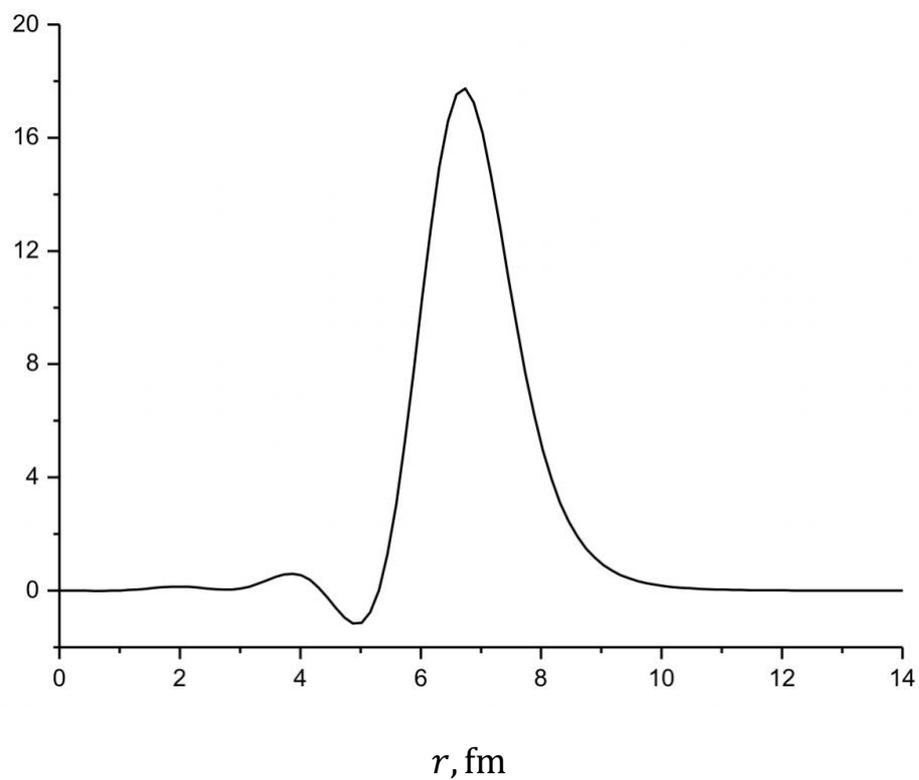

$r$, fm

**Fig. 8.** The projected transition density taken at the resonance peak-energy and calculated within PHDOM for High-Energy ISGOR (solid line), and the $3^-$ state transition density calculated within cRPA (thin line) for $^{208}$Pb.

$\rho_{V_{L=3}}(r, \omega_{L=3(peak)}), \text{fm}^{-1}\text{MeV}^{-1/2}$

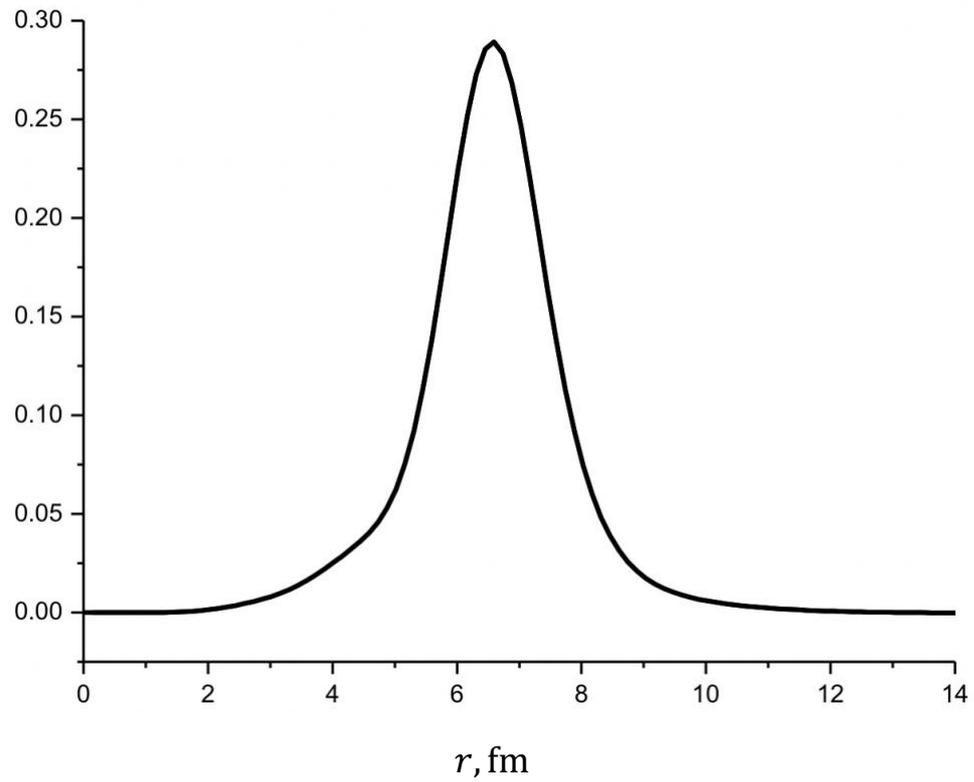

$r, \text{fm}$

$\rho_{L=3}^{coll}(r), \text{fm}^{-1}$

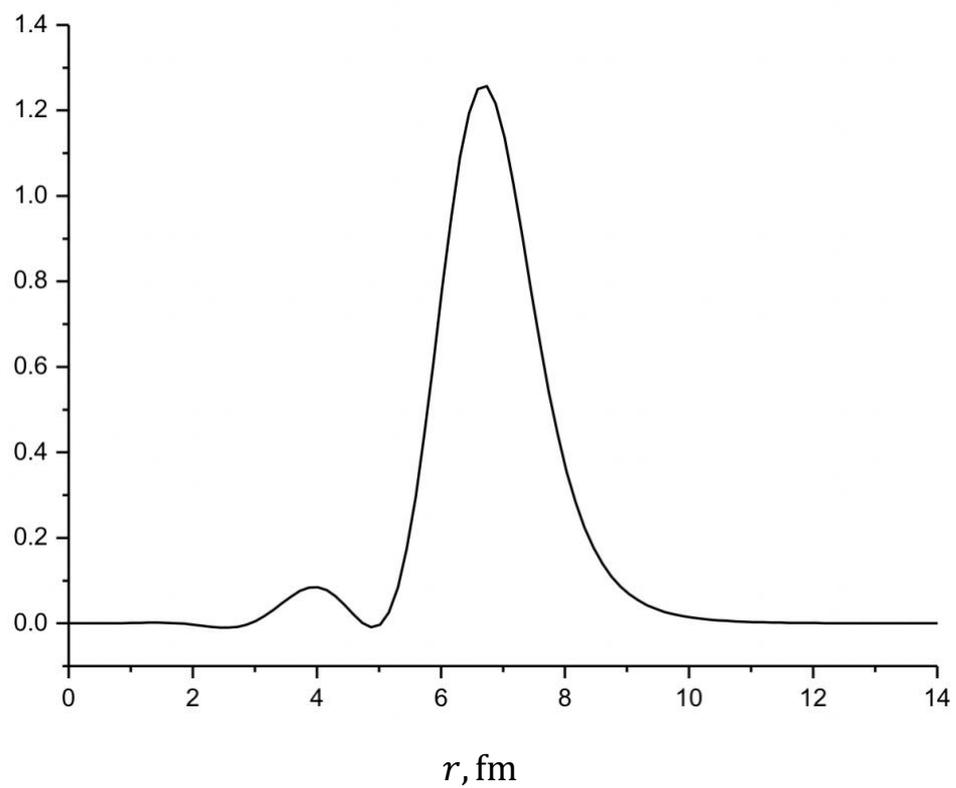

$r, \text{fm}$